\documentstyle[12pt]{article}
\newcommand{\be}{\begin{equation}}
\newcommand{\ee}{\end{equation}}
\newcommand{\bea}{\begin{eqnarray}}
\newcommand{\eea}{\end{eqnarray}}
\newfont{\doubleletter}{msbm10 scaled\magstep2}
\newfont{\unc}{eurb10 scaled\magstep 1}
\def\pd{\partial}
\makeatletter
\def\eqnarray{\stepcounter{equation}\let\@currentlabel=\theequation
\global\@eqnswtrue
\global\@eqcnt\z@\tabskip\@centering\let\\=\@eqncr
$$\halign to \displaywidth\bgroup\@eqnsel\hskip\@centering
  $\displaystyle\tabskip\z@{##}$&\global\@eqcnt\@ne
  \hfil$\displaystyle{\hbox{}##\hbox{}}$\hfil
  &\global\@eqcnt\tw@ $\displaystyle\tabskip\z@
  {##}$\hfil\tabskip\@centering&\llap{##}\tabskip\z@\cr}
\@addtoreset{equation}{section}
  \def\theequation{\thesection.\arabic{equation}}
\makeatother
\begin{document}
\begin{titlepage}
\date{\rightline {DTP/93/49}
\rightline{September 1993}}\title{Equations with infinitely many Lagrangians.}

\author{D.B. Fairlie.\\
 { \it University of Durham,}\\
 South Road, Durham, DH1 3LE, UK}

\maketitle

\begin{abstract}
Necessary conditions for a field theoretic equation of motion to be the
consequence of variation of an infinite number of inequivalent Lagrangians are
examined.
\end{abstract}
\end{titlepage}
\section{Introduction}
In this article, an avenue of investigation suggested by  the Universal Field
Equations proposed in \cite{fai} is explored.  Two of the striking properties
of these equations are that they possess an infinite number of inequivalent
Lagrangian formulations, and are completely integrable \cite{dbf}, a feature
which was anticipated by the presence of an infinite number of conservation
laws implied by the Lagrangian property. If, as is the case here the Lagrangian
does not depend explicitly upon the field, the Euler variational equation takes
the form of a conservation law. Also there exist many different conserved
stress tensors. These attributes suggest that they can be considered as a
halfway-house between ordinary Lagrangian and Topological Field Theories.
 The first two sections are devoted to an analysis of certain circumstances
under which an equation of motion for a scalar field, containing no
derivatives higher than the second in space-time of dimension two or greater
admits an infinite number of inequivalent Lagrangians. (Two Lagrangians which
differ by a divergence or are constant multiples of each other  are of course
equivalent.)
Hundreds of papers have been devoted to this topic, the inverse problem in the
calculus of variations, in the case where only time dependence is considered
\cite{mor}, but comparatively little has been done in the case of field
theories. It will be demonstrated that while not generic, the situation where
the inverse problem has a non-unique solution is not uncommon. However, if the
additional requirement of Lorentz or Euclidean invariance is imposed, the
possibilities are
rather limited, and this rules out most  such equations for any possible
application to physics. The final section establishes integrability
by the method of Legendre Transforms.

The Universal Field
Equation
\be
\det\pmatrix{0&{\pd \phi\over\pd x_k}\cr
               {\pd \phi\over\pd x_j}&{\pd^2 \phi\over\pd x_j\pd x_k}\cr}=0,\
i=1-d.
\label{1}
\ee
in $d$ dimensional spacetime of which the Bateman equation
\be
(\frac{\pd\phi}{\pd x})^2\frac{\pd^2\phi}{\pd t^2}+(\frac{\pd\phi}{\pd
t})^2\frac{\pd^2\phi}{\pd x^2}- 2\frac{\pd\phi}{\pd x}\frac{\pd\phi}{\pd
t}\frac{\pd^2\phi}{\pd x\pd t}=0.
\label{1a}
\ee
 is the two-dimensional case admits an infinite number of Lagrangians, the
construction of which is given in \cite{fai}. (for the Bateman case, any
Lagrangian density of the form
\be
{\cal L}= \frac{\pd\phi}{\pd t}F\Bigl(\frac{\frac{\pd\phi}{\pd
t}}{\frac{\pd\phi}{\pd x}}\Bigr)
\label{2d}
\ee
where $F$ is an arbitrary function will do)!

We shall argue that a necessary condition that the following generalisation of
(\ref{1});
\be
\sum_{i,j}G_{ij}\Bigl( \frac{\pd^2 \phi}{\pd x_k\pd x_l}\Bigr)_{ij}^{-1}=0.
\label{2a}
\ee
where $G$ is a $d\times d$ matrix with elements functions of first derivatives
$\frac{\pd\phi}{\pd x_k}$ and $\frac{\pd^2 \phi}{\pd x_k\pd x_l}$ the Hessian
matrix, regarded as an equation of motion, possesses an infinite number of
inequivalent Lagrangian formulations is that $\det G$ is of rank at most $d-1.$
In the case of (\ref{1}) the relevant matrix is of rank one, and this is
overkill! We shall demonstrate this, first by examining the
situation for $d=2$ and $d=3$, then by invoking a linearisation by means of a
Legendre transform suggested by \cite{dbf}, infer the general result.

\section{Infinitely many Lagrangians in 2 dimensions.}

Consider a scalar field theory with Lagrangian ${\cal L}(\phi_i),\ i=1,2$,
where $\phi_i$ denotes $\frac{\pd\phi}{\pd x_i}$.
The equation of motion is
\be
\sum_{i=1}^{i=2}\sum_{j=1}^{j=2}\frac{\pd^2{\cal L}}{\pd\phi_i\phi_j}\phi_{ij}=
G_{11}\phi_{11}-2G_{12}\phi_{12}+G_{22}\phi_{22}=0,
\label{2b}
\ee
in the notation of (\ref{2a}). From this formulation it is easy to make the
identifications
\be
\frac{\pd^2{\cal L}}{\pd\phi_1^2}=\lambda(\phi_i)G_{11};\
\frac{\pd^2{\cal L}}{\pd\phi_1\phi_2}=-\lambda(\phi_i)G_{12};\
 \frac{\pd^2{\cal L}}{\pd\phi_2^2}=\lambda(\phi_i)G_{22}.
\label{3a}
\ee
where a proportionality function $\lambda$ has been introduced which will be
determined up to a constant factor if $\cal L$ is unique.
Now if the equation of motion (\ref{2b}) is given, there are two integrability
constraints arising from (\ref{3a}) of the form
\bea
\frac{\pd\lambda}{\pd \phi_2}G_{11}+\frac{\pd\lambda}{\pd \phi_1}G_{12}+
\lambda(\frac{\pd G_{11}}{\pd \phi_2}+\frac{\pd G_{12}}{\pd \phi_1})&=0\\
\frac{\pd\lambda}{\pd \phi_2}G_{12}+\frac{\pd\lambda}{\pd \phi_1}G_{22}+
\lambda(\frac{\pd G_{12}}{\pd \phi_2}+\frac{\pd G_{22}}{\pd \phi_1})&=0,
\label{3d}
\eea
to be satisfied. In general, if these equations are linearly independent
they can be integrated to determine $\lambda$ uniquely in terms of the
functions $G_{ij}$
and their derivatives. If, instead, they are linearly dependent, then $\lambda$
retains a certain arbitrariness, which results in an arbitrariness in the
construction of $\cal L$ from (\ref{3a}). Necessary and sufficient conditions
for this eventuality are easily obtained as
\bea
\det\pmatrix{G_{11}&G_{12}\cr
             G_{12}&G_{22}\cr}=&0\\
\frac{\pd}{\pd \phi_1}\sqrt{\frac{G_{12}}{G_{11}}}-\frac{\pd}{\pd
\phi_2}\sqrt{\frac{G_{12}}{G_{22}}}=&0.
\eea
 These are evidently satisfied by the Bateman equation (\ref{1a}) for which
\be
G_{11}=\phi_2^2;\ G_{11}=\phi_1\phi_2;\ G_{22}=\phi_1^2.
\label{3c}
\ee
 More general solutions appear to be difficult to obtain, and in what follows
we shall restrict ourselves to the case where $g_{ij}$ is quadratic in
$\phi_{ij}$.
For the choice (\ref{3c}) the equation to determine $\lambda$ (\ref{3d})is
simply
\be
\phi_1\frac{\pd\lambda}{\pd \phi_1}+\phi_2\frac{\pd\lambda}{\pd
\phi_2}+3\lambda=0.
\ee
This simply means that $\lambda$ is a homogeneous function of $\phi_i$ of
degree
-3, a result which may be verified to be in accord with (\ref{2d}).

\section{Infinitely many Lagrangians in 3 and more dimensions.}

The starting assumption of the previous section must be modified to obtain a
nontrivial extension of these results to 3 dimensions. If the equation of
motion is simply linear in the second derivatives, with coefficients functions
of the first derivatives of the fields, then the constraints analogous to
(\ref{3d}) are overdetermined. Instead we start with a Lagrangian density of
the form
\be
{\cal L}=\sum_{i=1}^{i=3} \sum_{j=1}^{j=3} L^{ij}(\phi_k)\phi_{ij}
\label{4a}
\ee
Here the coefficients $L^{ij}(\phi_k)$ depend only upon the first derivatives
of
$\phi$ so that $\cal L$ depends only linearly upon the second derivatives of
the field. This is a redundant description, as three of these coefficients may
be removed by removing divergences to eliminate equivalent Lagrangians.
The equation of motion takes the form
\be
\sum_{i,j,k,l}
\frac{\pd^2L^{ij}}{\pd\phi_k\phi_l}(\phi_{ik}\phi_{jl}-\phi_{il}\phi_{jk})=0.
\ee
Now compare this equation with the equation of motion, assumed given in the
form
\be
\sum_{i,p,q}\sum_{j,r,s}\epsilon_{ipq}\epsilon_{jrs}G(\phi_k)_{ij}\phi_{pr}\phi_{qs}
\label{eqn}
\ee
Compatibility of those two equations requires the existence of a
proportionality factor $\lambda$ such that
\be
\sum_{p,q}\sum_{r,s}
\epsilon_{ipq}\epsilon_{jrs}\frac{\pd^2L^{pr}}{\pd\phi_q\phi_s}=\lambda G_{i.j}
\label{4b}
\ee
In this formalism it is easy to see that there are three equations of
consistency guaranteeing the integrability of (\ref{4b}), namely
\be
\sum_{i=1}^{i=3}\frac{\pd}{\pd \phi_i}(\lambda G_{ij})=0.
\label{4c}
\ee
The argument proceeds along the same lines as before; If the equations
(\ref{4c}) admit a solution for $\lambda$ which is not unique (discounting a
trivial proportionality constant) then a necessary condition is that $\det
G=0$,
i.e. the matrix of coefficients $G_{ij}$ must have rank 2 at most. There is
also a differential equation to be satisfied. At this point we introduce the
simplifying assumption that the $G's$ are quadratic in derivatives in order to
exhibit two solutions.
The first
\be
G_{ij}=\phi_i\phi_j
\ee
recovers the Universal Field Equation of \cite{fai} for the case of 3
dimensions. The matrix  $G$ is here of rank 1, a stronger requirement than
necessary. The general solution of (\ref{4c}) determines $\lambda$ to be an
arbitrary homogeneous function of $\phi_i$ of weight - 4.

The second is constructed using the following matrix representation for $G$;
\be
G=\pmatrix{\phi_2^2+\phi_3^2&\phi_3(\phi_1+\phi_2)&\phi_2(\phi_1+\phi_3)\cr
         \phi_3(\phi_1+\phi_2)&\phi_1^2+\phi_3^2&\phi_1(\phi_2+\phi_3)\cr
         \phi_2(\phi_1+\phi_3)&\phi_1(\phi_2+\phi_3)&\phi_1^2+\phi_2^2\cr}.
\ee
In this case $G$ is of rank 2, and the coefficients are such that the equations
(\ref{4c}) are equivalent to two independent equations;
\bea
\phi_1\frac{\pd\lambda}{\pd \phi_2}+\phi_2\frac{\pd\lambda}{\pd
\phi_3}+\phi_3\frac{\pd\lambda}{\pd \phi_1}+\lambda=0,\\
\phi_1\frac{\pd\lambda}{\pd \phi_3}+\phi_2\frac{\pd\lambda}{\pd
\phi_1}+\phi_3\frac{\pd\lambda}{\pd \phi_2}+\lambda=0.
\eea
These equations admit the symmetric solution, among many possible solutions,
 \be
\lambda=(\phi_1^2+\phi_2^2+\phi_3^2-\phi_1\phi_2-\phi_2\phi_3-\phi_3\phi_1)^{1+k}(\phi_1+\phi_2+\phi_3)^k,
\label{4f}
\ee
which depends upon a single parameter $k$. The Lagrangian $\cal L$ is then
determined up to divergences in terms of the three functions $L^{11},\ L^{22},\
L^{33}$ (setting the others to zero). For each choice of $k$ the equations to
for these functions may be solved, giving rise to an inequivalent set of
Lagrangian densities parametrised by $k$. Two specific cases are diplayed in
an appendix to this paper.

This example may be used to deduce an automatic construction of equations of
motion in $d$ dimensions which admit an infinite number of distinct Lagrangian
formulations.

Consider the following equation of motion;
\be
\sum_{i,j}G_{ij}{\rm Adj}\Bigl( \frac{\pd^2 \phi}{\pd x_k\pd x_l}\Bigr)_{ij}=0,
\label{4k}
\ee
where
\be
{\rm Adj}\Bigl( \frac{\pd^2 \phi}{\pd x_k\pd x_l}\Bigr)_{ij}=\det\Bigl(
\frac{\pd^2 \phi}{\pd x_k\pd x_l}\Bigr)\Bigl( \frac{\pd^2 \phi}{\pd x_k\pd
x_l}\Bigr)_{ij}^{-1}
\ee
is the adjugate matrix, and $G_{ij}$ is constructed in the following manner.
Take $\phi_i^P=P((\phi_i)$ to be a permutation of the components of the vector
$\{\phi_1,\phi_2,\dots\phi_d\}$ by an element $P$ of the permutation group
in $d$ dimensions. Then  take $G_{ij}$ to be a linear sum  with arbitrary
constant coefficients of terms of the form
$\phi_i^P\phi_j^P$, taken over subset of not more than $d-1$ elements $P$
of the permutation group. This will guarantee that $\det G$ is of rank at most
$d-1$. The case which corresponds to the three dimensional situation above is
obtained by taking a sum over cyclic permutations. Take a differential operator
representation for $P_j$ as
\be
P_j=\phi_1\frac{\pd}{\pd \phi_{j+1}}+\phi_2\frac{\pd}{\pd \phi_{j+2}}+\cdots+
\phi_d\frac{\pd}{\pd \phi_{j}}
\ee
 where the indices are defined in an obvious cyclic manner and take
\be
G_{ij}=P_1(\phi_i)P_1(\phi_j)+P_2(\phi_i)P_2((\phi_j)+\cdots+P_{d-1}(\phi_i)P_{d-1}(\phi_j).
\ee
Then, introducing $\lambda$ in an obvious generalisation of the 3 dimensional
situation above the $d-1$ consistency equations to determine $\lambda$  become
\be
P_n\lambda+\lambda=0,\   n=1,2,\dots,d-1.
\label{4h}
\ee
Suppose $\omega_j$ is a $d$th complex root of unity and we define $\omega_0=1.$
Then the function
$f_j=\sum_{i=1}^d\omega_j^{i-1}\phi_i$ is an eigenfunction of the differential
operator $P_1$ with eigenvalue $\omega_j^{d-1}=\omega_j^*$. Then a solution to
the equations (\ref{4h}) dependent upon an arbitrary parameter $k$
is evidently given by
\be
\lambda=\prod_{j=1}^{j=d-1}f_j^{k+1}f_0^k
\ee
in analogy with (\ref{4f}). If $d$ is composite, then there is an obvious way
to include additional parameters in this type of solution, by multiplication by
arbitrary powers of other zero eigenfunctions. For each such $\lambda$, the
equations to determine a Lagrangian will be integrable, and thus an infinite
number of Lagrangians, parametrised by $k$ will exist. Equations of this sort,
unlike the Universal Equation, have lost Lorentz invariance and retain only
discrete transformations among the space-time variables. This means that as far
as physics is concerned, such equations can be discounted.

\section{Legendre Transforms}
  Further insight into the circumstances under which an infinite number of
Lagrangian descriptions of a theory exist is afforded by the Legendre
Transform, which was used in \cite{fab} to linearize the Universal Field
Equation.  This transform, which is clearly involutive has the flavour of a
twistor transform.
and also works for equations of the type (\ref{4k}).
The multivariable version of this transform runs as follows \cite{cour}
Introduce a dual space with co-ordinates $\xi_i,\ i=1,\dots,d$ and a function
$Q(\xi_i)$ defined as follows
\bea
\phi(x_1,x_2,\dots,x_d)+Q(\xi_1,\xi_2,\dots,\xi_d)=
&x_1\xi_1+x_2\xi_2+\dots ,x_d\xi_d.\\
\xi_i={\pd\phi\over\pd x_i},\quad x_i={\pd Q\over\pd \xi_i},\quad& \forall{i}.
\eea
To evaluate the second derivatives $\phi_{ij}$ in terms of derivatives of $Q$
it is convenient to introduce two Hessian matrices;
$\Phi,\ {\cal Q}$ with matrix elements  $\phi_{ij}$ and ${\cal
Q}_{\xi_i\xi_j}=q_{ij} $
respectively. Then assuming that $\Phi$ is invertible, $\Phi {\cal Q}=1\!\!1$
and
\be
 {\pd\sp2\phi\over\pd x_i\pd x_j}= ({\cal Q}\sp{-1})_{ij},\quad
 {\pd\sp2q\over\pd \xi_i\pd \xi_j}= (\Phi\sp{-1})_{ij}.
\ee

 The effect of the Legendre transformation upon the equation (\ref{4k}) is
immediate; in the new variables the equation becomes simply
\be
\sum_{i,j}G_{ij}(\xi_k) \frac{\pd^2 Q}{\pd \xi_i\pd \xi_j}=0,
\label{4l}
\ee
a linear second order equation for $q$. In the case discussed above, with a
permutation operator representation for $P_j$ now as
\be
P_j=\xi_1\frac{\pd}{\pd \xi_{j+1}}+\xi_2\frac{\pd}{\pd \xi_{j+2}}+\cdots+
\xi_d\frac{\pd}{\pd \xi_j}
\ee
then equation (\ref{4l}) becomes
\be
\sum_{i=1}^{i=d-1}\sum_{j=1}^{j=d-1}(P_iP_j- P_{i+j})Q=0.
\ee
Now the crucial point is that this equation does not affect the dependence of
the function $Q$ upon the particular combination represented by a zero
eigenfunction of all $P_j$, i.e.
\be
\prod_{j=0}^{j=d-1}(\sum_{i=1}^{i=d} \omega_{j}^{i-1}\xi_i)
\ee. This means that it is effectively an equation in a manifold of dimension
$d-1$ instead of $d$. Thus the phenomenon of the existence of an infinite
number of Lagrangians in the original space, is related to the fact that the
linear differential operator in the transform space acts only upon a subspace
of the transform variables .

\section{Appendix: Examples}
Two examples of Lagrangians for the equation (\ref{eqn}) are as follows:
Corresponding to parameter $k=-1$

\be
{\cal L} = -\sum_{i=1}^{3}\phi_{ii} \phi_{i} (\phi_{i} \Phi\  log(\Phi)
-\phi_{i+1} \phi_{i+2})
\ee
where
\[\Phi=\phi_{1}+\phi_{2}+\phi_{3} \]
and the indices are defined cyclically. The Lagrangian corresponding  to
parameter $k=0$ is given by

\bea
{\cal L} &=&
\phi_{11} (-5 (\phi_{2}^6+\phi_{3}^6-2 \phi_{1}^6)+9 \phi_{1}^5
(\phi_{2}+\phi_{3}) +9 \phi_{1} ( \phi_{3}^5+\phi_{2}^5) \nonumber \\
&+&
45 \phi_{2} \phi_{1} \phi_{3} (\phi_{2}^3 +\phi_{3}^3)-15 ( \phi_{2}^2
\phi_{1}^4+\phi_{2}^4 \phi_{1}^2+\phi_{3}^2
\phi_{1}^4+\phi_{1}^2 \phi_{3}^4) \nonumber \\
&+&
30 \phi_{2} \phi_{1}^2 \phi_{3} (3 (\phi_{2} \phi_{1}-\phi_{2}
\phi_{3}+\phi_{1} \phi_{3})-2  \phi_{2}^2 +
\phi_{3}^2)))\\
&+& {\rm cyclic\ replacements.} \nonumber
\eea

\end{document}